\documentclass[12pt]{article}

\usepackage{amscd,amsfonts,amssymb,amsthm,amscd,amsmath}
\usepackage{bm}
\usepackage{caption}
\usepackage[mathscr]{eucal} 
 \usepackage{latexsym}
\usepackage{mathptmx}
\usepackage{mathrsfs} 
\usepackage{mathtools}

\usepackage{color}
\usepackage{cite}
\usepackage{epsfig}
\usepackage{graphicx}

\numberwithin{equation}{section}

\def\cA{{\cal A}}
\def\cB{{\cal B}}

\def\cG{{\cal G}}
\def\cH{{\cal H}}

\def\cL{{\cal L}}
\def\cM{{\cal M}}

\def\cO{{\cal O}}
\def\cP{{\cal P}}

\def\cS{{\cal S}}

\def\BH{{\cB(\cH)}}

\def\fA{{\mathfrak A}}

\def\fR{{\mathfrak R}}

\def\CC{{\mathbb C}}

\def\NN{{\mathbb N}}

\def\RR{{\mathbb R}}

\def\ZZ{{\mathbb Z}}

\def\ovac{\omega_{\, 0}}
\def\Hvac{\cH_{\, 0}}
\def\pivac{\pi_{\, 0}}
\def\vac{\Omega_0}        
\def\Uvac{U_0} %

\newcommand{\bF}{{\bm F}}
\newcommand{\bR}{{\bm R}}

\newcommand{\bU}{{\bm U}}

\newcommand{\bZ}{{\bm Z}}

\newcommand{\bGamma}{{\pmb \Gamma}}

\newcommand{\Pid}{{\cP_+^\uparrow}}
\newcommand{\Lid}{{\cL_+^\uparrow}}

\def\eg{\textit{e.g.\ }}

\def\ie{\textit{i.e.\ }}
\def\viz{\textit{viz.\ }}

\newcommand{\ad}[1]{\mbox{ad}  \, #1}

\newcommand{\be}{\begin{equation}}
\newcommand{\ee}{\end{equation}}

\begin{document}

\title{Algebraic quantum field theory: 
objectives, methods, and results}

\author{\large Detlev Buchholz${}^{(1)}$ \ and 
\ Klaus Fredenhagen${}^{(2)}$ \\[5mm]
\small 
${}^{(1)}$ Mathematisches Institut, Universit\"at G\"ottingen, \\
\small Bunsenstr.\ 3-5, 37073 G\"ottingen, Germany\\[5pt]
\small
${}^{(2)}$ 
II. Institut f\"ur Theoretische Physik, Universit\"at Hamburg \\
\small Luruper Chaussee 149, 22761 Hamburg, Germany \\
}
\date{}

\maketitle

\begin{abstract}
  \noindent Algebraic quantum field theory is a
general mathematical framework for relativistic quantum physics, based on the
theory of operator algebras. It comprises all observable
and operational aspects of a theory. In its framework the entire state space
of a theory is covered, starting from the vacuum over arbitrary
configurations of particles to thermal equilibrium and
non-equilibrium states. It provides a solid foundation for
structural analysis, the physical interpretation of the theory and the
development of new constructive schemes.
This  survey is commissioned by the {\it Encyclopedia of Mathematical
Physics}, edited by M.~Bojowald and R.J.~Szabo. It is to be 
published by the Elsevier publishing house. 
\end{abstract}

\section{Origin and achievements}
\label{sec1}
\setcounter{equation}{0} 
Algebraic quantum field theory (AQFT) emerged from the framework  
of quantum field theory \cite{StWi,e1},
which relies on the principle of locality
(the quantum version of Maxwell's Nahwirkungsprinzip). 
The primary goal of quantum field theory is the
description of relativistic particles and their interactions. But
there appeared several questions.
\begin{itemize}
\item[(a)] Non-interacting field theories were
  known to describe particles that 
  propagate freely. But how can one extract from a theory 
  its particle aspects in the presence of interaction? 
\item[(b)] Quantum physics exhibits non-local phenomena, such as 
  entanglement and the violation of Bell inequalities. How is this
  compatible with the locality principle, in particular with the 
  maximum signal velocity of light?
\item[(c)] It was realized that there are theories 
  with different field content and 
  different Lagrangians which describe the same physics.
  Can one identify physically indistinguishable theories by some
  meaningful equivalence relation? 
\item[(d)] Are there properties of quantum field theories that are generic, \ie
  independent of specific models?
\end{itemize}

Algebraic Quantum Field Theory answered these and related 
questions of physical interest in a general framework, based on first
principles. It arose from the problem of computing a 
scattering matrix for the 
multitude of elementary particles and their compounds
that were observed in high energy experiments. The idea to 
assign to each of these interacting particles a separate
quantum field, as in non-interacting theories, seemed to be odd.  
This problem was solved by Haag \cite[II.4]{Ha}. He observed that it
is sufficient for the computation of scattering matrices 
to exhibit for each one-particle state suitable 
operators, built out of a few fundamental fields,
which have non-vanishing matrix elements
between the vacuum and the one-particle state. There
exist many operators with this property in general, but the resulting
scattering matrices do not depend on their specific choice. 

Based on this insight, Haag proposed to reformulate quantum field
theory as follows \cite[III]{Ha}:
instead of dealing with the technically subtle
point-localized quantum fields, 
one takes as basic input the algebras of genuine operators generated by 
fields in bounded subregions of spacetime. Thereby one focuses
on combinations of the fields describing observables. This restriction 
allows one to postulate properties of
direct physical significance for the resulting algebras.
As a matter of fact, it turned out to be irrelevant that the algebras are
generated by quantum fields.
 
The fundamental paradigm is that the linking of algebras with 
spacetime regions uniquely characterizes a theory.  
The mathematical framework of AQFT is put on a rigorous basis
by the Haag-Kastler
axioms \cite{HaKa}, which rely on Einstein causality and
the Poincar\'{e} symmetry of Minkowski space.
Later it was extended to generally covariant theories on
globally hyperbolic spacetimes. Here we restrict our attention
to theories on four-dimensional Minkowski space.

A crucial feature of AQFT, describing systems with an infinite
number of degrees of freedom, is the existence of disjoint representations
of the algebras by Hilbert space operators. They describe macroscopically
distinguishable systems differing, for example, by temperature, global
charges, or their behavior at infinity.  
This fact required an overview of the  representations, in
particular the identification of representations describing elementary states,
such as the vacuum and single particle states. 

Vacuum representations are characterized by the existence of a
Poincar\'e invariant ground state; these representations may also
describe single particle states, such as the photon.
But charged single particle 
states do not belong to vacuum representations of the observables.
The corresponding charged representations were
characterized by Doplicher, Haag, 
and Roberts \cite{DoHaRo1} by the property that they cannot be distinguished
from the vacuum representation by observables
which are localized in the spacelike complement of given
bounded regions. This excludes states with electrical
charge because of Gauss' law, but includes states with nonzero
baryon or lepton number.
We say, these states carry a localizable charge.
With this input, Doplicher, Haag and Roberts
were able to completely unravel the structure
of the corresponding representations.
They clarified the origins of Bose and Fermi 
statistics, the existence of global gauge groups that describe the
charges of particles, and they constructed charged
Bose and Fermi fields that interpolate between the vacuum and
the charged states \cite{DoHaRo1,DoHaRo2,DoRo}.
The dimension of physical spacetime played an
important role in this context. In two spacetime dimensions other 
forms of statistics and symmetries can occur.  

The analysis of gauge theories, in particular of the
confinement problem, made it clear that it is not sufficient to
consider only particles with localizable charges. 
Another approach, taken by Buchholz and Fredenhagen \cite{BuFr1}, therefore
started directly with the discussion of representations,
where the bottom of the energy-momentum spectrum contains some
isolated mass hyperboloid, characterizing the states of a single
particle. There then exists an associated
vacuum representation which coincides with the particle representation
on observables localized in the \mbox{complement} of cone-shaped
regions extending to spacelike infinity. Hence in massive
theories all particles can be localized in such cones.
These weaker localizability properties still allowed it to
arrive at results similar to those obtained by Doplicher, Haag and
Roberts for localizable charges. Other types of statistics
and symmetries can appear for cone-localizable states
already in three spacetime dimensions.

In theories describing long range forces, such as quantum electrodynamics, 
the charged particle structure is less well understood. The reason is that
the particle states always contain 
infinite clouds of low energy massless particles. For them a
meaningful classification is not available. However, in physical
spacetime one can make use of the fact that 
massless particles propagate with the
speed of light (Huygens principle).
Thus the probability of finding an unlimited number
of them in a given forward lightcone,
is equal to zero. 
One therefore restricts the 
states to the observables in lightcones with fixed apex 
and concentrates on the properties of the resulting 
representations. In case of charged states carrying a simple
charge, like the electric one, one arrives in this manner
at similar results as in massive theories. In particular, the
charge structures are described by abelian gauge groups~\cite{BuRo}.
 
There are other states of great physical interest
which describe thermal equilibria.
Since in infinite space Hamiltonians of physical interest
have continuous spectrum,
the Gibbs-von Neumann description of equilibrium states
by density matrices in the vacuum representation can not be used,
the thermal representations are disjoint from the vacuum representation. 
Instead, one characterizes the equilibrium states by specific
analyticity properties of their correlation functions,
the so-called KMS condition, named after Kubo, Martin, and
Schwinger and invented by Haag, Hugenholtz, and Winnink
\cite{HaHuWi,e2}.
In order to identify those theories which admit equilibrium states for
all positive temperatures, one needs criteria which restrict the
size of state space \cite{HaSw,BuWi}. These criteria 
impose constraints on the nature of correlations between
observations in spacelike separated regions.
As a matter of fact, the correlations can be completely
suppressed in suitable states \cite{Bu,BuDaFr}.
This feature allows it to construct
equilibrium states in finite spacetime regions which coincide
in their spacelike complements with the vacuum.
Thinking of high energy physics, it resembles the formation of
a quark-gluon plasma. Enlarging the regions occupied by these
localized equilibrium states, they have thermodynamic limits 
which satisfy the KMS condition \cite{BuJu}.
 
There is a surprising relationship  between the KMS condition in
algebraic quantum field theory and a cornerstone
in the theory of operator algebras, \viz modular theory, invented
by Tomita and Takesaki \cite{Ta,e3}. In the mathematical theory one studies 
suitable algebras of operators on a Hilbert space, which do
not contain elements (apart from $0$) that annihilate a given state.  
It turns out that these algebras have an intrinsic time evolution for which the
correlation functions of the given state satisfy the KMS condition.
Thermal states in AQFT fit into this framework. But 
the mathematical results also imply, for example, that the vacuum
state is a KMS states with regard to the intrinsic dynamics 
of the algebra of observables localized in a wedge region, bounded
by two lightlike planes. The intrinsic time evolution is in
this case the one-parameter group of Lorentz boosts, which
leaves the wedge invariant. In physical terms, uniformly
accelerated observers register in the vacuum state some
non-zero temperature (Unruh effect) \cite{Se}. 

The connection of AQFT with modular theory led to a variety
of fruitful applications that could not be achieved within the
conventional formalism of quantum fields. It helped to
establish duality relations (Haag-duality) 
between observables in spacelike separated
regions, which is crucial for the analysis of
the properties of superselection
sectors \cite{BiWi}. It was also used in arguments establishing 
the universal structure of local algebras \cite{Fr2}.
A remarkable property of these algebras is the absence of
finite dimensional projections. This must be kept in mind
when discussing measurements and operations. 
Arguments that claim their apparent
acausal behavior often ignore this fact and are therefore 
invalid. As a matter of fact, modular theory has been
a key ingredient in the
discussion of entanglement.
Last, but not least, it
has become an important tool in the rigorous constructions of
models which are not accessible by other methods, such as theories
with factorizing scattering matrices \cite{Le,e4}. 
 
These results confirmed the conviction that the physical content of
a theory is encoded in the net structure of the underlying local 
algebras of observables. The question of how to characterize within
the general formalism a specific theory (\eg by the 
\mbox{reconstruction} of a corresponding Lagrangian function
from the algebras) remained open, however. 
Only recently it has turned out that a complementary view 
on this issue is more fruitful \cite{BuFr2}. Regarding fields and their
composites, such as Lagrangian functions, as classical objects,
corresponding quantum operations can rigorously be defined.
It provides for given Lagrangian function, involving local interactions, 
a corresponding net of local algebras satisfying the 
basic postulates of AQFT. The scheme works for theories involving
Bose as well as Fermi fields and leads, for example,
to an algebraic understanding of \mbox{renormalization} and related
issues, such as the appearance of anomalies\cite{BrDuFrRe1,BrDuFrRe2}.
The prospects for a
general constructive scheme within AQFT that applies to all theories
of physical interest, including gauge theories,  are promising.

In the following the underlying notions, specific
methods, and key results are outlined in greater detail.
References to articles in this encyclopedia have
an E in front of their number. Other references
are given by their numbers only.

\section{The algebraic framework}
\label{sec2}
\setcounter{equation}{0} 

As in any quantum theory, the observables and resulting basic operations 
are described in AQFT by elements $A$ of some non-commutative associative
algebra $\fA$ over the complex numbers $\CC$ together with an antilinear
involution $A \mapsto A^*$. The algebra $\fA$
may be thought of as being concretely given in some
defining representation by
operators acting on a Hilbert space, where $A^*$ is the adjoint of
$A$. But, as an abstract algebra, 
it has a multitude of other representations by Hilbert space operators. 
One usually refers to $\fA$ as algebra of observables.
 
Observables correspond to selfadjoint elements $A =  A^* \in \fA$,
and basic operations are described by unitaries $U \in \fA$.
Their adjoint action $\ad{U}(A) \coloneqq UAU^*$ on
any observable $A$ does not 
change its spectrum and corresponding multiplicities. 
Assuming that the spectrum of the observables $A$ 
is bounded, which can always be accomplished by a
suitable (non-linear) choice of scale, 
they have a bounded norm $\| A \|$, inherited from the
defining Hilbert space representation.
This norm satisfies for any $A \in \fA$ the condition
$\| A^* A \| = \| A \|^2$, the so-called C*-property.
Completing $\fA$ with respect to this norm, one 
obtains a C*-algebra ~\cite[Sect.\ III.2.1]{Ha}.
It corresponds to the norm closure of a subalgebra of
the algebra of all bounded operators in the defining representation.
 
It is a  distinctive property of AQFT that it comprises information
as to where and when measurements are made,
\ie about their localization in Minkowski space $\cM$.
(The metric on $\cM$, used here, is positive on
timelike vectors, the velocity of light is $c = 1$.) 
Thus, given any bounded spacetime region $\cO \subset \cM$, 
there is a corresponding subalgebra $\fA(\cO) \subset \fA$ 
containing all operators which correspond to 
measurements or operations within the ranges of $\cO$.
As these sets of operators become bigger if the localization region
increases, one has the inclusions 
\be \label{e.2.1}
\fA(\cO_1) \subset \fA(\cO_2) \quad \mbox{if} \quad 
\cO_1 \subset \cO_2 \, .
\ee
Due to this property of isotony, the
mapping of spacetime regions
to algebras, $\cO \mapsto  \fA(\cO)$, constitutes a 
net based on Minkowski space $\cM$ from which  
the algebra $\fA$ of all observables can be recovered 
in the limit $\cO \nearrow \cM$. The principle
of Einstein causality implies that observations in spacelike
separated regions must be commensurable. Whence, because of the 
finite propagation speed of physical effects, they 
cannot disturb each other in any way. This basic fact is encoded in the
condition of locality, 
\be \label{e.2.2}
[ \fA(\cO_1) , \, \fA(\cO_2)] = \{0\}  \quad \mbox{if} \quad 
\cO_1 \perp \cO_2 \, .
\ee
It says that the commutator of any pair of observables which are  
localized in spacelike separated regions,
indicated by $\cO_1 \perp \cO_2$,  
has to vanish.

Relativity enters by assuming that the
symmetry group of Minkowski space $\cM$, 
the proper orthochronous Poincar\'e group $\Pid = \Lid \ltimes \RR^4$,
acts by automorphisms $\alpha(\lambda$) on
$\fA$, $\lambda \in \Pid$. Recalling that the
elements of $\Pid$ relate inertial
observers to each other, the action of the automorphisms on
the local algebras satisfies   
\be  \label{e.2.3} 
\alpha(\lambda)(\fA(\cO)) = \fA(\lambda \cO) \, , \quad \lambda \in \Pid
\, ,
\ee
in an obvious notation. Thus the localization regions of the algebras 
change under the automorphisms 
according to the geometric action of the corresponding
Poincar\`e transformations on Minkowski space.  
These physically meaningful conditions define a mathematical 
framework for the observables and operations in any physically acceptable
relativistic quantum theory on Minkowski space \cite[III]{Ha}.
It provides a basis for
their general analysis, their physical interpretation,
and it paves the way for the construction of models. 


\section{Algebraic constructions}
\label{sec3}
\setcounter{equation}{0} 

At present, the existence of interacting quantum fields
in physical spacetime has been accomplished only by perturbation
theory, \ie field operators are defined in terms of
formal power series whose convergence properties are not
under control. These methods have been refined and transferred
to the algebraic setting, leading to perturbative AQFT
\cite{BrDuFr,Re,e5}.
Quite recently, the insights gained in these investigations led
to a new constructive scheme \cite{BuFr2}. It yields for given classical
Lagrangian function a concrete algebra which complies with all
postulates of AQFT. This construction is outlined
in the following for scalar selfinteracting quantum fields.

The ingredients in this approach are arbitrary classical,
real, and smooth scalar fields $x \mapsto \phi(x)$
on Minkowski space $\cM$, which may be unbounded
at infinity. Corresponding classical local 
observables are described by functionals of the form 
\be \label{e.3.1}
\phi \mapsto F[\phi] \coloneqq \sum_{n=0}^N \int \! dx \, f_n(x)
\phi(x)^n \, ,
\ee 
where $x \mapsto f_n(x)$ are real test functions with compact support
in $\cM$. The
support of a functional $F$ in Minkowski space is defined as the 
union of the supports of the underlying test functions $f_n$,
$n = 1, \dots , N$. The functional for $n=0$ has empty
support. It can be assigned to any given spacetime region. 

Lagrangian functions (densities) on $\cM$, describing
local self-interactions of the classical field, are of the form
\be \label{e.3.2}
x \mapsto L(x)[\phi] \coloneqq (1/2) \, (\partial_\mu \phi(x) \,
\partial^\mu \phi(x) - m^2 \phi(x)^2)
- \sum_{j = 1}^J g_j \, \phi(x)^j \, ,
\ee
where $\partial_\mu$ denotes the partial derivative with regard to the
$\mu$-coordinate of $x$, $m$ is some mass value, and $g_j$ are real
coupling constants. Integrating these Lagrangian 
functions over all of $\cM$, whenever meaningful for
a given field~$\phi$, yields a value of its action.
 
However, the integral defining the action will in
general not converge. But local variations of the action can always be defined. 
To this end one introduces shifts of 
the functionals $F$, putting
\mbox{$F^{\phi_0}[\phi] \coloneqq F[\phi + \phi_0]$},
where $x \mapsto \phi_0(x)$ are arbitrary real and smooth
scalar fields 
with compact support. Applying these shifts to the Lagrangian
functions, one finds that 
$x \mapsto (L(x)[\phi + \phi_0] - L(x)[\phi])$ 
can be integrated over all of $\cM$ for any field $\phi$. 
Thus the variations of the corresponding actions given by
\be \label{e.3.3}
\delta L(\phi_0)[\phi]
\coloneqq \int \! dx \, (L(x)[\phi + \phi_0] - L(x)[\phi])
\ee
are well defined for all fields $\phi$.
As a matter of fact, one finds by partial integration that $\delta L(\phi_0)$
is a functional of the form \eqref{e.3.1} for fixed 
$\phi_0$.

Based on this classical input, one defines for given
Lagrangian function~$L$ a corresponding dynamical group $\cG_L$ 
that is associated with AQFT. It is the
group generated by elements $S_L(F)$, satisfying two basic
relations, where $F$ are arbitrary
functionals as defined in equation~\eqref{e.3.1}. 

 \noindent 
(1) \ Identifying the
localization of $S_L(F)$ in Minkowski space with the support
of the corresponding functional $F$, a first equality 
describes causal relations between these elements. 
Whenever the support of a functional $F_1$ is later than
that of a functional $F_2$, \ie there is some Cauchy surface
in $\cM$ such that $F_1$ lies above and $F_2$ beneath it, one has
\be \label{e.3.4}
S_L(F_1) S_L(F_2) = S_L(F_1 + F_2) \, .
\ee
Note that the product on the left hand side is not
commutative unless the supports of the functionals are
spacelike separated; only then can one find
Cauchy surfaces separating the supports in
either temporal order. Hence relation \eqref{e.3.4} is an expression
of relativistic causality which is more stringent than
the condition of locality. This relation also implies that
the constant functionals $\phi \mapsto c[\phi] \coloneqq c$,
$c \in \RR$, determine elements $S_L(c)$ of the center of
$\cG_L$. Moreover, they satisfy the equation 
$S_L(c_1) S_L(c_2) = S_L(c_1 + c_2)$.
\\[2mm]
(2) \ The second relation, involving the dynamical input determined
by the Lagrangian function $L$, is given by \
\be \label{e.3.5}
S_L(F) = S_L(F^{\phi_0} + \delta L(\phi_0)) 
\ee
for all functionals $F$ and shift fields $\phi_0$. This
equality is an integrated version of the
Schwinger-Dyson equation for quantum fields, where 
the given Lagrangian function $L$ enters \cite{BuFr2,e5}. 
 
Heuristically, the elements $S_L(F)$ 
can be interpreted as basic operations which describe the
impact of perturbations induced by $F$ on the underlying
systems. They may be thought of as time ordered exponentials
of $(i/\hbar)F$  acting on a quantum field.
But no explicit quantization procedures are used 
in their construction. As a matter of fact, the dynamical group $\cG_L$
is non-commutative from the outset due to the arrow
of time that enters in the defining relation \eqref{e.3.4}. 

We want to interpret the operations $S_L(F)$ as
  unitary operators acting on Hilbert spaces. To this end
  we put $S_L(c) \coloneqq e^{\, ic} \, 1$ for the constant
  functionals $c \in \RR$, which is compatible with
  relations \eqref{e.3.4} and \eqref{e.3.5}.
  It amounts to choosing atomic units, 
  where Planck's constant is $\hbar = 1$. A simple example of 
  such a Hilbert space with an action of the operations
  can be constructed as follows. The space
  is generated by the complex linear span of vectors  
  $|S \rangle$, $S \in \cG_L$, 
  where $|S_L(c) S \rangle \coloneqq e^{ic} |S \rangle$. 
  The scalar product for these generating vectors is given by 
\be
\langle S|S'\rangle \coloneqq 
\begin{cases}
        e^{ic} & \text{if} \ \  S'=S_L(c)S = e^{ic} S \\
            0 & \text{otherwise.} 
\end{cases} 
\ee
The action of the group $\cG_L$ on this space
is defined by $S|S'\rangle \coloneqq |SS'\rangle$. Thus the 
elements $S \in \cG_L$ act as invertible isometric operators and hence
are unitary.

All linear combinations of
these unitaries, which by the distributive law for products 
form an algebra $\fA_L$, are faithfully represented on this space,
\ie do not vanish unless they are identically $0$. To see this, consider
the sum $\sum_k c_k S_k$, where group elements differing only
by some phase factor are combined in a single term with an
appropriate c-number factor. Applying this operator to the
vector~$| 1 \rangle$, the resulting vector has the norm-square
$ \sum_k |c_k|^2$ and hence vanishes only if all coefficients $c_k$ 
are equal to~$0$, as claimed. So the operator norm 
on the Hilbert space determines a C*-norm on~$\fA_L$.
In a similar manner, any other faithful Hilbert space representation
of $\fA_L$ determines a norm on it. Since the C*-norm of any unitary operator
is equal to~$1$, the supremum of all C*-norms on $\fA_L$ is well defined
and one can equip the algebra with this maximal C*-norm and complete
it in this topology. The result is a C*-algebra, which is denoted by the same
symbol.

The local subalgebras
$\fA_L(\cO) \subset \fA_L$ are generated by 
elements $S_L(F)$, where $F$ 
has support in $\cO \subset \cM$. So the net
$\cO \mapsto \fA_L(\cO)$ satisfies the condition of
locality. There exist also automorphisms on the net which
induce Poincar\'e transformations $\Pid$. This follows from the fact that the
underlying classical Lagrangian functions transform as scalar
fields under their action. Hence $\fA_L$ satisfies
all Haag-Kastler axioms.

This approach not only leads to a rigorous construction of 
theories fitting into the general framework of AQFT for
a large set of Lagrangians, but
it also provides a basis for computations. We briefly
illustrate this fact in the simple case of a non-interacting  Lagrangian
$L_0$, where all coupling constants in equation~\eqref{e.3.2}
are put equal to $0$. One considers for arbitrary real
test functions $f$ on $\cM$ the functionals $F_W(f)$ of the form
\be \label{e.3.7}
\phi \mapsto F_W(f)[\phi] \coloneqq
(1/2) \int \! dx dy \, f(x) \Delta_D(x-y) f(y)
+ \int \! dz \, f(z) \phi(z) \, ,
\ee
where $\Delta_D = (1/2) \, (\Delta_R + \Delta_A) $ is the mean of
the retarded and advanced solutions
of the Klein-Gordon equation with mass $m$. The first
term on the right hand side of~\eqref{e.3.7} 
defines some constant functional, the second one is
linear in the underlying field.

Putting
$W(f) \coloneqq S_{L_0}(F_W(f))$ and making use of relations
\eqref{e.3.4} and \eqref{e.3.5} yields after some computations 
the equalities \cite{BuFr2} 
\begin{align}  \label{e.3.8}
& W(f_1) W(f_2) = e^{-(i/2) \int \! dx dy \,
  f_1(x) \Delta(x-y) f_2(y)} \, W(f_1 + f_2) \, , \nonumber \\
& W((\square + m^2) f_3) = 1 \, .
\end{align}
Here $f_1, f_2, f_3$ are real test functions,
$\square$ is the d'Alembertian, and 
\mbox{$\Delta \coloneqq (\Delta_R - \Delta_A)$} (Pauli-Jordan function).
Thus the operators $W(f)$ are unitary exponentials of
a real, scalar, local quantum field that satisfies the Klein-Gordon
equation and is integrated with test functions $f$ (Weyl operators).
These well known operators are elements of the algebra~$\fA_{L_0}$
and appear in the present approach without imposing any commutation relations
from the outset. The exponent of the phase factor reveals
the use of atomic units, where $\hbar = 1$. 

We conclude this outline  by noting that a more refined
version of the causal\-ity relation \eqref{e.3.4}, involving
higher products, has further interesting consequences~\cite{BuFr2}. 
For example, given any Lagrangian $L$ of the form \eqref{e.3.2},
the corresponding net of local algebras $\cO \mapsto \fA_L(\cO)$ consists 
of subalgebras of the global algebra $\fA_{L_0}$, determined by
the non-interacting Lagrangian $L_0$; but these
subalgebras differ from
the local algebras in the non-interacting theory. In other words,
different theories merely differ by the resulting nets
in a fixed global algebra. In this way one arrives
at an algebraic version of the interaction picture.   
Vacuum representations for different Lagrangians are,
however, inequivalent in agreement with Haag's Theorem.
   
The algebraic constructions, explained here in simple cases,
have been extended to theories involving an arbitrary finite
number of interacting bosonic and fermionic fields. Moreover,
several questions of physical interest have been settled. 
Among them are the time evolution of given initial data 
(time slice axiom), the occurrence of symmetries (Noether's
theorem), and the impact of changes of renormalization
(renormalization group) \cite{BrDuFrRe1,BrDuFrRe2,e5}.

This new constructive approach has contributed to the consolidation
of AQFT, showing that C*-algebras satisfying the Haag Kastler
axioms exist in presence of interaction. An important 
future step will be the analysis of their state spaces. 
Given a dynamical algebra, one may be able to show 
that it has a vacuum state, in analogy to  
the existence proofs of constructive
quantum field theory. Or one may be able to show that
for a given dynamics the corresponding algebra 
does not have any such state, cf.\ \cite{Ai,Fro,e6}. 

\section{States and representations}
\label{sec4}
\setcounter{equation}{0}

The algebras $\fA$, complying with the basic assumptions
of AQFT, are designed to explore the properties of 
all ensembles which can appear in a theory.
Dealing with quantum theory, one makes statistical 
predictions about measuring results. These are 
encoded in expectation functionals 
$\omega : \fA \rightarrow  \CC$, \ie linear maps from
the elements of the algebra to complex 
numbers. Each ensemble in which a system can be 
prepared determines some functional $\omega$. 
Given $A \in \fA$, the entity $\omega(A)$ is interpreted
as expectation value (mean of measuring results) in the 
corresponding ensemble. Since the variance of an 
observable cannot be negative 
and $A^*A$ is an observable with
non-negative spectrum, one demands $\omega(A^*A) \geq 0$ for  
$A \in \fA$. One also requires the normalization 
$\omega(1) = 1$. Any such positive and normalized functional $\omega$
on~$\fA$ is called a state. 
 
It is an important fact that every state 
$\omega$ determines a representation of 
the algebra $\fA$, the GNS representation~\cite[Sect.\ III.2]{Ha},
which acts by operators on some Hilbert space. 
It is given by a triple 
$(\pi, \cH, \Omega)$ consisting of a Hilbert space $\cH$,
a unit vector $\Omega \in \cH$ representing the given 
state, and a structure preserving map
(homomorphism) $\pi : \fA \rightarrow \BH$ 
of the algebra $\fA$ into the algebra $\BH$ of 
bounded operators on $\cH$. The link between the 
two settings is provided by the formula
\be \label{e.4.1}  
\omega(A) = \langle \Omega,  \pi(A) \Omega \rangle \, , \quad 
A \in \fA \, .
\ee
Thus expectation functionals of elements of $\fA$ are represented by 
matrix elements of corresponding operators on $\cH$. 
Given a state $\omega$, the corresponding GNS-representation is unique,
up to unitary equivalence. 
Dealing with infinite systems, different states in a
theory lead in general to quite different (inequivalent) Hilbert
space representations. Hence the usage of the concept of
state is more flexible than starting 
with some particular Hilbert space representation of the
algebra. 

Given a state $\omega$, one obtains other states by two physically
meaningful operations. The first one corresponds to perturbations,
which are caused by measurements or by 
operations such as local changes of the dynamics. 
They are described by 
operators $V \in \fA$ which are scaled such that
$\omega(V^* V) = 1$. The resulting states are given by
$A \mapsto \omega_V(A) \coloneqq \omega(V^*AV)$, $A \in \fA$.
The second operation is the formation of 
mixtures of these perturbed states, which are described by
convex combinations.
The norm closure of the resulting convex set of states
is called the folium of $\omega$. All states in  this folium can be described
by density matrices in the GNS--representation induced by~$\omega$.

Two states are said to be disjoint if their respective folia 
have an empty intersection. In that case there exist classical
observables distinguishing the folia. They are obtained by 
sequences of observables whose commutators with any other
observable tend to $0$ and whose expectation values
converge to different numbers in the two folia.
Prominent examples are global charges, temperature, and order parameters
distinguishing different phases. The situation is particularly simple in the
case of pure states, \ie states which cannot be decomposed into convex
combinations of other states. Pure states describe ensembles  with maximal
information and induce irreducible representations.
The folia of any two pure states are either
disjoint or they coincide. One therefore speaks of sectors
of the entire state space which are formed by these folia.
 
Let us finally mention some important technical point in this context. 
If a sequence 
$\{A_n \in \fA \}_{n \in \NN}$ converges in norm to some operator $A$,
  this entails the uniform convergence of its expectation values, 
\be
\lim_n \, \sup_{\omega} |\omega(A_n-A)| =0 \, ,
\ee
where the supremum is taken over all states. 
In a given folium one can consider a weaker form of convergence.
One demands that for any state in the folium,
the expectation values converge to those of 
some bounded operator $A$, which may
not be contained in $\fA$, however. Completing the local algebras in this
weak topology yields von Neumann algebras (also called W*-algebras)
in the representation fixed by the folium.  
These completions contain for example the projection
operators appearing in 
the spectral decomposition of local observables,
which enter in the probabilistic interpretation of the
underlying states. Different folia
induce in general different weak topologies, so it
is not meaningful to work with this topology from the
outset. 


\section{Elementary states}
\label{sec5}
\setcounter{equation}{0}

On any C*-algebra $\fA$ there exists an abundance of states and
corresponding representations. Many of them are of limited 
physical interest since they describe over-idealizations, such
as infinite accumulations of matter.
It is therefore of importance to identify those
states which are essential for the physical interpretation
of the theory. 

Conceptually, the simplest states are
vacuum states \cite[III.4]{Ha}. A vacuum state $\ovac$ 
(there may be several such states or none) 
is by definition a ground state which looks alike for all inertial observers.
It therefore is invariant under Poincar\'e transformations, 
$\ovac \, \alpha(\lambda) = \ovac$, $\lambda \in \Pid$,
where the product of the state and the Poincar\'e automorphism
indicates their composition. This implies that there exists 
in the GNS-representation $(\pivac, \Hvac, \vac)$, 
induced by  $\ovac$, a unitary representation $\Uvac$ of
the Poinar\'e group which is given by, $A \in \fA$,  
\be \label{e.5.1} 
\Uvac(\lambda) \, \pivac(A) \, \vac \coloneqq \pivac(\alpha(\lambda)(A)) \, \vac
\, , \quad \lambda \in \Pid \, .
\ee
The correlation functions
$\lambda \mapsto \ovac(A^*\alpha(\lambda)(A))$
are supposed to be continuous, \mbox{$A \in \fA$}, so the unitary representation
$\lambda \mapsto \Uvac(\lambda)$ is weakly continuous.
Hence the subgroup of unitaries representing the spacetime translations
$x \in \RR^4$ can be presented by Stone's theorem in the form
$\Uvac(x) = e^{ixP}$ with generators $P$, which are interpreted
as energy-momentum operators. Their joint spectrum can be determined
by Fourier analysis of the correlation functions.
Since a vacuum state is a ground state for all inertial observers,
the spectrum must be contained in the forward lightcone
\mbox{$V_+ = \{ p \in \RR^4:  p_0 \geq 0, \, p^2 \geq 0 \}$},
where $p_0$ denotes  
the energy component and $p^2$ the Minkowski square of $p$. 
Finally, vacuum states can always be 
uniquely decomposed into a convex combination of disjoint pure vacuum states.
We restrict our attention in the following to pure vacuum states
  and assume that the corresponding GNS-representations are faithful,
  \ie they are regarded as defining representation of $\fA$. 

Vacuum states have many interesting properties. Among
them are clustering properties of vacuum correlation functions
\cite{ArHeRu},
the absence of local operators annihilating the
vacuum (Reeh-Schlieder property \cite{ReSch}), the entanglement of
spacelike separated operations \cite{SuWe,e7}, and the energetic
effects of the spontaneous breakdown of internal symmetries
(algebraic Goldstone theorem \cite{BuDoLoRo}). Because of lack of
space, most of these results cannot be discussed here in detail. 

The folium of the vacuum state contains only states which are
neutral in the sense that they can be obtained by applying observables
to the vacuum vector, a prominent example being the photon.
Charged particles lead, by definition, to disjoint representations
of the underlying algebra. 

The determination of the charged particle
content of a theory is based on two physical 
ideas. The first one, going back to 
Haag and Kastler \cite{HaKa}, is
to consider sequences of states in the folium of the vacuum
that consist, heuristically, of a charge in a fixed region
and a compensating charge in another region which is
moved to spacelike infinity.   
The compensating charge at infinity does then no longer contribute  to
the expectation values,
so the limit states carry only the original charge. The second idea 
has been advocated by Borchers \cite{Bo}  and is based on the
condition that elementary states of interest should
be pure and admit a 
representation of the group of spacetime translations 
with generators having joint spectrum in~$V_+$.
This spectrum is automatically invariant under Lorentz transformations
as consequence of locality \cite{BoBu}.  
In theories where only massive particles appear, one
can identify charged single particle states by the
condition that the spectrum does not contain the
discrete point $p=0$, corresponding to the
vacuum, but an isolated mass shell $p^2 = m^2$ for some
mass $m>0$. As will be discussed, this input can be used for the
construction and analysis of composite states containing
several particles.

This approach 
fails, however, in theories with long
range forces, such as quantum electrodynamics.
There particle states carrying an electric charge
can neither be localized in bounded regions of Minkowski
space because of the Coulomb field which they carry along. Nor
is their mass shell separated from the rest of the
spectrum due to clouds of low energy photons which inevitably
accompany them as a consequence of Gauss's law and locality. 
There is progress in the identification of such particles
\cite{BuPoSt,e8,e9}, but this topic deserves further studies.  


\section{Sectors, statistics, and charged fields}
\label{sec6}
\setcounter{equation}{0}

Having identified the elementary charged states in a theory,
a number of subsequent questions arise. First, given two
charged states, does there exist a state containing both
charges (addition of charges)? Second, does there exist
for each charged state a state carrying the opposite
charge (charge conjugation)? Third, can one assign to charged
states a particular statistics (Bose-Fermi alternative)?
And forth, do there always exist charged fields which
create the charged states from the vacuum
and transform as tensors under the action of some global gauge group?
These questions require an answer, since one has initially
only the observable algebra $\fA$ of a theory at ones disposal and
charge-carrying fields are not given from the outset.

For localizable charges, all of these questions have found an
affirmative answer in AQFT. 
Starting from the
characterization of elemen\-tary states as local excitations
of the vacuum, Doplicher, Haag and Roberts
have established these facts in extensive investigations
\cite{DoHaRo1,DoHaRo2,DoRo}.
These were later supplemented by Buchholz and Fredenhagen
for theories containing exclusively 
massive particles \cite{BuFr1}. The latter results
cover theories, where certain particle states fail to be 
localized excitations of the vacuum since they carry gauge
or topological charges. The central ideas and methods underlying
these results are outlined in the following. 

In order to characterize elementary states $\omega$
on a given algebra $\fA$ which are \mbox{local} excitations
of a vacuum state $\ovac$, one compares the respective 
expectation values of observables that are localized in distant regions. 
Given a causally closed
region~$\cO$, which has the form of a double-cone
(intersection of a forward and a backward lightcone), let
$\cO^c$ be its spacelike complement.
The corresponding subalgebra
$\fA(\cO^c) \subset \fA$  is defined  as the algebra
generated by all local observables having their support in
double-cones contained in $\cO^c$. The state $\omega$ is then said to
describe a local excitation of $\ovac$
if for every $\varepsilon > 0$ there exists
some sufficiently large double cone $\cO$ such that 
\be \label{e.6.1}
\| \omega - \ovac \|_{\, \cO^c} \coloneqq
\sup_{A \in \fA(\cO^c) \, , \ \|A\| = 1} |\omega(A) - \ovac(A)| <
\varepsilon \, .
\ee
In simple terms, the two states cannot be distinguished 
by measurements at large spacelike distances. 

It is a consequence of 
relation \eqref{e.6.1} that the 
states $\omega$ and $\ovac$, restricted to $ \fA(\cO^c)$,
induce equivalent representations of this subalgebra  
even though they are disjoint on $\fA$. Moreover, the GNS
representation of $\fA$ induced by $\omega$ is faithful.
Using methods of the theory of operator algebras, 
it follows that it can be represented on the vacuum Hilbert
space $\cH_0$ in a manner such that it coincides with the vacuum
representation on $ \fA(\cO^c)$.  This representation
can be continued to the weak closures of the local
algebras in the vacuum representation. For  
the sake of simple notation, we denote in this section
the weak closures of $\pivac(\cA(\cO))$ by
$\fR(\cO) \coloneqq \pivac(\cA(\cO))^-$ and
the inductive norm limit of the resulting net
$\cO \mapsto \fR(\cO)$ by $\fR$.
We also assume that the algebras $\fR(\cO)$ are maximal
in the sense that they contain every operator which
commutes with all
operators $A^c \in \pivac(\cA(\cO^c))$ (``Haag duality'').
The representation of 
the extended algebra $\fR \supset \pivac(\fA)$, induced by $\omega$,
is denoted by $(\rho, \cH_0, \Omega_0)$, that is 
\be \label{e.6.2} 
\omega(R) = \langle \Omega_0, \rho(R) \, \Omega_0 \rangle \, ,
\quad R \in \fR \, .
\ee
Recalling that the GNS-representations are unique up to 
unitary equivalence, the vector $\Omega_0 \in \cH_0$, which
represents $\ovac$ in the vacuum representation, now
represents the charged state $\omega$ in the representation $\rho$.  
The localization properties of
$\omega$ and the feature of Haag duality imply that the
represented algebra $\rho(\fR)$ is
not just any subalgebra of bounded operators on $\cH_0$: 
it is contained in the
original domain $\fR$ of $\rho$. Moreover, the representation
$\rho$ coincides with the vacuum representation on the algebra
$\fR(\cO^c)$, 
\be \label{e.6.3}
\rho(R) = R  \quad \mbox{if} \quad  R \in \fR(\cO^c) \, ,
\ee
revealing the fact that the charge can be localized in the
bounded region $\cO$.

In view of these properties, the first question raised above has 
an immediate answer: given representations $\rho_1$ and $\rho_2$
that are induced by elementary charged states, a representation
containing both charges is obtained by their composition~$\rho_1 \rho_2$.
This composition is well defined since the ranges
of the representations are contained in their domains $\fR$. The answer to
the second question, concerning the existence of opposite charges,
is immediate for the family of representations~$\rho$
induced by states carrying so-called simple charges
\cite{DoHaRo1}. The corresponding 
representations $\rho$ are distinguished by the property that they map
the algebra $\fR$ onto itself, so they have an inverse
$\rho^{-1}$. An example of a simple charge 
is univalence, which distinguishes Bosons from Fermions.
 The representation $\rho^{-1}$ is localized in the
same region as $\rho$ and compensates the charge of $\rho$
by composition. Only these simple charges are discussed in the following;
an outline of the  general case of non-simple charges,
where $\rho(\fR)$ is a proper subalgebra of $\fR$,
would require more space.

Coming to the third question concerning statistics, one makes use of
the fact that the considered representations are translation covariant,
so translations of the charges do not change their values.
To see this, let us recall that the translations in the
vacuum representation act by unitaries $U_0$ 
on the vacuum Hilbert space $\cH_0$; their adjoint action
on $\fR$, denoted by $\alpha_0$, leaves this algebra invariant.
Similarly, the translations in representations $\rho$
induced by elementary states act by unitaries $U_\rho$ on $\cH_0$,
hence their adjoint action is defined on  $\fR$ as well. 
Given a representation $\rho$ with charge contained in $\cO$,
as explained above, the translated representation
with charge contained in $\cO + x$ is given by 
\be \label{e.6.4} 
    {}^x \rho \coloneqq \alpha_0(x) \, \rho \, \alpha_0(-x)
    = \ad{(U_0(x) U_\rho(-x))} \, \rho
    \, , \quad
   x \in \RR^4 \, .
\ee 
Thus the representations $\rho$ and $ {}^x \rho$ are unitarily
equivalent, whence they carry the same charge. The unitaries
(called charge transporters)
which connect the representations are elements of the 
local algebra $\fR(\cO_{0,x})$, where $\cO_{0,x}$ is any 
double cone containing $\cO$ and $\cO + x$. So they 
are elements of the algebra $\fR$ for
all localizable charges. It is noteworthy that 
the adjoint action of the unitaries
$U_\rho(x) U_0(-x)$ on~$\fR$ allows one to
recover in the vacuum sector the charged
representation $\rho$ in the limit of large
spacelike $x \in \RR^4$. This procedure corresponds to
the heuristic idea of creating charged states from bi-localized
neutral states.

The statistics of a charged representation $\rho$ is determined
by analyzing the products (compositions) ${}^x\rho {}^y \rho$ for
translations $x,y \in \RR^4$. One first notices that the resulting
representations are all equivalent to the representation
$\rho^2 \coloneqq \rho \rho$ since the charge-transport operators
are elements of $\fR$. In the case of simple charges, considered
here, $\rho^2$ maps $\fR$ onto itself again. One then considers
the products for translations
$x,y$ such that the charge of ${}^x \rho$ in $\cO +x$ 
is spacelike separated from the charge  of ${}^y \rho$
in $\cO + y$. It is an important consequence of the locality 
property of the observables that 
${}^x\rho {}^y \rho = {}^y\rho {}^x \rho$, \ie the creation of
charges in spacelike separated regions does not depend on their
order \cite{DoHaRo1}. Putting $\Gamma(x) \coloneqq U_0(x) U_\rho(-x)$, 
one has 
\be \label{e.6.5}
   {}^x\rho {}^y \rho = \ad{(\Gamma(x))} \, \rho \, \ad{(\Gamma(y))} \rho
   = \ad{( \Gamma(x) \rho(\Gamma(y) )} \, \rho^2 \, ,
   \quad x,y \in \RR^4 \, ,
\ee
since the composition of maps is associative. 
Now $\rho^2(\fR) = \fR$ is irreducibly represented in the
representation induced by the pure vacuum state. Hence
relation \eqref{e.6.5} 
implies that the unitary charge-transport operators
$\Gamma(x) \rho(\Gamma(y))$, respectively  $\Gamma(y) \rho(\Gamma(x))$, 
shifting the
two charges in $\rho^2$ into spacelike separated regions
$\cO +x$, respectively $\cO + y$, can only differ by some phase
factor~$\epsilon_\rho(x,y)$,
\be \label{e.6.6}
\Gamma(x) \rho(\Gamma(y)) = \epsilon_\rho(x,y) \Gamma(y) \rho(\Gamma(x)) \, ,
\quad \cO + x \perp \cO + y \, .
\ee
In a final deformation argument one uses the fact that
in four spacetime dimensions one can exchange continuously any two
spacelike separated double cones whilst keeping them spacelike
apart. This fact entails that the  
phase factors do not depend on the admissible translations,
\ie one has $\epsilon_\rho \coloneqq \epsilon_\rho(x,y)$ for all such
$x,y$. Moreover, one obtains for their  square $\epsilon_\rho^2 = 1$. Thus there
exists for each sector of a simple
charge, described by a representation $\rho$,
a corresponding unique number $\epsilon_\rho \in \{ \pm 1 \}$,
called statistics parameter.

The statistics parameters enter in the 
commutation relations of field operators 
that render the representations $\rho$.  We briefly indicate their 
construction for representations describing
self-conjugate charges; there the representation $\rho^2$
is equivalent to the vacuum representation,
\ie $\rho^2 = \ad{V}$ for some unitary operator
$V \in \fR(\cO)$. In this case the field operators are 
defined on the direct sum of the neutral and 
the charged representation space, described by 
the vectors $(\Phi, \, \Psi)$, where $\Phi, \Psi \in \cH_0$.
The observables and translations are
represented there by 
\be \label{e.6.7}
\bR \coloneqq 
\begin{pmatrix}
  R & 0 \\
  0 & \rho(R) 
\end{pmatrix} \! ,
\ \, R \in \fR \, , \qquad
\bU(x) \coloneqq
\begin{pmatrix}
  U_0(x) & 0 \\
  0 & U_\rho(x)
\end{pmatrix} \! ,
\ \, x \in \RR^4 \, .
\ee
One then defines unitary field operators $\bF$ on this space, putting
\be  \label{e.6.8}
\bF \coloneqq 
\begin{pmatrix}
  0 & 1 \\
  V & 0
\end{pmatrix} \, ,
\qquad
\bF^* = 
\begin{pmatrix}
  0 & V^* \\
  1 & 0
\end{pmatrix} \, ,
\ee
where $V \in \fR(\cO)$ is the unitary operator given above.
This yields, in an obvious notation, 
\be \label{e.6.9}
\bF \bR \bF^* = \begin{pmatrix}
  \rho(R)  & 0 \\
  0 & VRV^*
\end{pmatrix} = \begin{pmatrix}
  \rho(R)  & 0 \\
  0 & \rho^2(R)
\end{pmatrix}
= \rho(\bR) \, ,  \quad R \in \fR \, .
\ee
Hence the field implements the action of $\rho$
and commutes with all observables in~$\fR(\cO^c)$.
In this sense, it is localized in $\cO$. 
Putting $\bF(x) \coloneqq \bU(x) \bF \bU(x)^*$ for $x \in \RR^4$,
it follows that
\be \label{e.6.10} 
\bGamma(x) \coloneqq \bF(x) \bF^* = \begin{pmatrix}
  U_0(x) U_\rho(x)^* &  \hspace{-3mm}  0 \\
  0 & \hspace{-3mm} U_\rho(x) V U_0(x) V^*
\end{pmatrix}
  = \begin{pmatrix} \Gamma(x) & \hspace{-3mm}  0 \\
  0 & \hspace{-3mm} \rho(\Gamma(x))  
\end{pmatrix}  \! .
\ee
Thus $\bGamma(x)$ comprises the charge transfer operators of 
the representations $\rho$, respectively
$\rho^2 = \ad{V}$.
The properties of these transfer operators, established
in relation \eqref{e.6.6}, lead to the equality
for all $x,y$ such that $\cO + x \perp \cO + y$
\be \label{e.6.11}
\bF(x) \bF(y) = \bGamma(x) \rho(\bGamma(y)) \bF^2
= \epsilon_\rho \, \bGamma(y) \rho(\bGamma(x)) \bF^2
= \epsilon_\rho  \, \bF(y) \bF(x) \, .
\ee
So, depending on the sign of $\epsilon_\rho$, the field operators
localized at spacelike distances satisfy Bose, respectively Fermi,
commutation relations. Which sign appears is encoded in the
net structure of the underlying observable algebra.

Let us finally remark
that there exists a unitary representation of the group
$\ZZ_2$ on the underlying representation space which is given by
\be \label{e.6.12}
\bZ = \bZ^* = \begin{pmatrix} 1 & 0 \\
  0 & -1   
\end{pmatrix}  \! .
\ee
Its adjoint action leaves all observables $\bR$ invariant and
changes the sign of the fields $\bF$, so it is an 
example of a unitary representation of a
global gauge group. To summarize, proceeding in the
present example from 
two disjoint elementary states on the algebra of observables,
one obtains an extension of the observable algebra 
by local field operators, satisfying either Bose or
Fermi commutation relations at spacelike distances.
They generate from the vacuum state, together with the observables,
a Hilbert space containing all charged and neutral states.
Using them, one can then compute collision states
and scattering matrices \cite{DoHaRo2,e8}, such as in
standard quantum field theory.
However, in contrast to the standard setting, the present
approach takes into account all particles occurring in a theory, 
including those for which no fields were initially specified. 
This is essential for proofs of asymptotic
completeness \cite{e8}.

After this survey of methods underlying sector analysis, let us
summarize the wealth of results which have been obtained
on this topic so far.
First, these investigations were performed for all kinds
of localized elementary states \cite{DoHaRo2}. The resulting representations
$\rho$ then no longer need to have an inverse;
they may be morphisms which map the observable
algebra into itself, not onto.
The products of these morphisms can generically be
decomposed into finite direct sums of morphisms which
induce irreducible representations of the observables
on the underlying vacuum Hilbert space. 
Every such irreducible morphism determines a
statistics parameter that indicates its Bose,
respectively Fermi (para)statistics.
Moreover for any such 
morphism $\rho$ there exists a conjugate morphism
$\overline{\rho}$ whose product with $\rho$ contains the
vacuum representation $\iota$.

The construction of
field operators in this general framework turned 
out to be difficult and required advanced methods from
category theory. It was finally accomplished by Doplicher
and Roberts \cite{DoRo}. They proved that for
the collection of localized morphisms in a theory
there exists a unique extension of the observable algebra
by compactly localized field operators, resulting in 
a field algebra. That algebra generates 
from the vacuum a representation space of the observables,
describing all finite configurations
of charged and neutral elementary systems. 
On this space, there acts a faithful unitary representation of a
compact group under whose action the fields transform as tensors,
while the observables remain fixed.
Thus this group operates as a global gauge group.
The fields satisfy Bose, respectively Fermi commutation
relations at spacelike distances. Finally, there exists
on the representation space a continuous
unitary representation of the covering group of the
Poincar\'e group with positive energy. Its adjoint
action on the fields transforms their localization
regions covariantly, in accordance with their
statistics. So also in this general case
the structures have been established which were taken for
granted in standard quantum field theory. 

There remained, however, the question whether this
analysis covers all situations of physical interest in theories
with short range forces, describing exclusively massive particles
at large scales. This question was raised and answered
by Buchholz and Fredenhagen \cite{BuFr1}. Without
any \textit{a priori} assumption about localization
properties, they proceeded from an elementary charged state
on the algebra of
observables. If the bottom of the energy-momentum spectrum
in the resulting representation consists of    
some isolated mass shell, signaling a
single particle, there exists also an accompanying  
vacuum state. The states in the sector of the particle are
excitations of the vacuum which can be localized in spacelike 
cones $\cS \subset \cM$ that extend to spacelike infinity. 
More precisely, relation~\eqref{e.6.1} holds for the norm
distance between the two states if the region~$\cO^c$
is replaced by $\cS^c$, the spacelike complement
of any spacelike cone $\cS$ containing some
sufficiently big neighborhood $\cO$ of the origin.

Given such a particle state and spacelike 
cone $\cS$, one obtains as in  
case of localizable charges a representation
of the algebra of observables 
which acts on the vacuum Hilbert space and is
identical to the vacuum
representation on the algebra $\fA(\cS^c)$. 
But in general it does not map the full algebra of observables
into itself, so defining the composition of these  
representations requires more work.
Remarkably, all basic results obtained for localizable
charges can also be established in this  
case. Among them are the Bose, respectively Fermi,
(para)statistics of sectors, the existence of conjugate sectors, 
of a field algebra generated by charged field operators satisfying Bose,
respectively Fermi commutation relations, and
of a global compact gauge group acting on the
fields and leaving the observables invariant.
Moreover, the possibility of infinite statistics,
which was originally left open in case of
localizable charges, was proven not to occur
in massive particle theories \cite{Fr1}.
These results cover massive gauge theories, where 
non-confined charged particles appear at asymptotic time,
or particles carrying a fluctuating topological charge. 

As already mentioned, a complete understanding of the sector structure
and statistics has not yet been accomplished for theories describing long
range forces between local observables. The reasons are the infinite
clouds of massless particles which are created by these interactions. 
They lead to an abundance of disjoint infrared sectors which cannot
be discriminated in experiments. In other words, the theoretical concept
of sectors is too subtle in these cases. In order to isolate the 
physically relevant structures, Buchholz and Roberts proposed to form
equivalence classes of sectors, making use of the fact that
infrared sectors cannot be distinguished by local observables in any
given lightcone \cite{BuRo}. This feature fits with the fact that one
cannot make up measurements in the past, and it explains why infrared
sectors do not play a role in experiments.

Based on these insights, Buchholz and Roberts 
performed an investigation of the sector structure of the
algebras of observables localized in any given
\mbox{lightcone}. Restricting
attention to sectors which are excitations of a vacuum state
in spacelike (hyper)cones, they succeeded in identifying the sectors
carrying simple charges. They then established the fact that these
sectors have the same properties with regard to statistics, charge
conjugation, existence of a global gauge group, and of charged fields
as the sectors in massive theories in Minkowski space. These
results complement in some sense an investigation of localizable charges
in arbitrary globally hyperbolic spacetimes \cite{GuLoRoVe,e10}. It would
be desirable to extend it to all elementary cone-localized
systems in lightcones in order to determine their possible 
structures. 

We conclude this section by noting that investigations
of the sector structure have also been performed in low spacetime
dimensions. There other types of statistics can appear,
described by representations of the braid group. The sector structure
can then in general no longer be described by the representation
theory of a compact group, it is determined by other
group-like structures. We refer the interested reader to \cite{e11}.


\section{Thermal states and modular theory}
\label{sec7}
\setcounter{equation}{0}

Elementary systems play a fundamental role in the interpretation
of the microscopic properties of a theory. One then aims to extract
from it its macroscopic features, such as the properties of
thermal equilibrium states. Conversely,
the requirement that a theory ought to have a decent macroscopic
behavior leads to constraints on its microscopic structure.
We outline in this section some basic \mbox{results} in this respect
and indicate a few consequences of physical interest. 
A more extensive discussion of equilibrium states can be found in 
\cite{e2}. 

In the standard approach to equilibrium states one proceeds from
finitely extended systems, \eg confined in a box, and considers
Gibbs-von Neumann ensembles whose density matrix is
described by exponential functions of the negative Hamiltonian,
multiplied with the inverse of the temperature $T > 0$. This works
whenever the level density of the system is such that the
trace of this operator is finite. That feature disappears,
however, if one
proceeds to the thermodynamic limit. Dealing in AQFT from the
outset with infinitely extended systems, there arises the
question whether one can recover from it this level density,
which encodes information about the number of degrees of 
freedom in a theory.

The idea for such a procedure goes back to Haag and Swieca \cite{HaSw};
it was later refined by Buchholz and Wichmann \cite{BuWi}.
One considers excitations of
the vacuum state $\omega_0$ on the observable algebra $\fA$
which are localized
in given bounded spacetime regions $\cO$ and suppresses their 
high energy contributions. Thereby, one obtains
in the resulting GNS-representation linear
maps $\theta_{T, \cO} : \fA(\cO) \rightarrow \cH_0$, 
\be \label{e.7.1}
\theta_{T, \cO}(A) \coloneqq e^{ - (1/T) H } \pi_0(A) \Omega_0 \, ,
\quad A \in \fA(\cO) \, ,
\ee
where $H$ is the generator of the time translations in a chosen 
Lorentz frame. \mbox{Restricting} these maps to operators in the unit ball
of $\fA(\cO)$, one determines their ranges in the vacuum Hilbert
space $\cH_0$. Heuristically, one considers the smallest Hilbert-box 
into which such a range fits, \ie
an infinite dimensional cuboid in $\cH_0$, 
centered at $0$, which contains it. Of primary interest are theories,
where the sums of the side lengths of these cuboids are finite;
these sums replace the partition functions of finite systems. One can then 
define corresponding norms $\| \theta_{T, \cO} \|_1$ of the maps.
If these norms are finite, the maps are said to be nuclear.

Theories, where these norms exhibit a physically meaningful behavior
in the limit of large temperatures and regions, have been
shown to admit thermal
equilibrium states on the algebra of observables
for all temperatures \cite{BuJu}. An important
intermediate step in the proof is the demonstration that these
theories have the so-called split property \cite{BuDaFr,DoLo}: given any pair
of bounded regions $\cO_1 \Subset \cO_2$, \viz the closure
of $\cO_1$ is contained in the interior of $\cO_2$,
there exists a vector $\Omega_{\cO_1,\cO_2} \in \cH_0$ such that
for all operators $R_1 \in \pi_0(\fA(\cO_1))$ and
$R_2' \in \pi_0(\fA(\cO_2))'$ one has
\be  \label{e.7.2}
\langle \Omega_{\cO_1,\cO_2}, R_1 R_2' \, \Omega_{\cO_1,\cO_2} \rangle
= \langle \Omega_0, R_1 \Omega_0 \rangle 
\langle \Omega_0, R_2'  \Omega_0 \rangle  \, .
\ee
Here the prime ${}'$ indicates the
algebra of all bounded operators on $\cH_0$ which commute
with the given algebra. By locality,
$ \pi_0(\fA(\cO_2))' \supset \pi_0(\fA(\cO_2^{\, c})$,  
so the equality shows that there exist product states in the
vacuum sector in which measurements in
a bounded region and its spacelike separated complement
are completely uncorrelated. One then considers
the projection $E_{\cO_1, \cO_2}$ onto the subspace of $\cH_0$  
spanned by 
\be  \label{e.7.3}
R_1 \, \Omega_{\cO_1,\cO_2} \, , \quad R_1 \in  \pi_0(\fA(\cO_1)) \, .
\ee
This projection commutes with all elements of $\pi_0(\fA(\cO_1))$
and transfers the elements of $R_2' \in \pi_0(\fA(\cO_2))'$
into the vacuum state,
\be \label{e.7.4} 
E_{\cO_1, \cO_2} \, R_2' \, E_{\cO_1, \cO_2} =
\langle \Omega_0, R_2' \Omega_0 \rangle \, E_{\cO_1, \cO_2} \, , \quad
R_2' \in \pi_0(\fA(\cO_2))' \, .
\ee
As an aside, if one replaces in \eqref{e.7.3}
the algebra $\pi_0(\fA(\cO_1))$ by $\pi_0(\fA(\cO_2))'$,
one arrives at a projection in $\pi_0(\fA(\cO_2))^-$
that transfers the elements of $\pi_0(\fA(\cO_1))$
into the vacuum state. This entails an algebraic  
version of the split property. 
Turning back to the problem at hand, one finds that
the operators $E_{\cO_1, \cO_2} e^{-(1/T) H} E_{\cO_1, \cO_2}$ have
a finite trace on $\cH_0$. Dividing the operators by its
value, one
obtains density matrices which replace in the present setting the  
Gibbs-von Neumann ensembles. Moreover, for suitably increasing
regions $\cO_1 \Subset \cO_2$ approaching $\cM$, the
resulting sequences of states have weak limits on the
algebra $\fA$, which are designed to describe global
equilibrium states for the given temperature $T$.

As was shown by Haag, Hugenholtz, and Winnink, 
it is a distinctive property of equilibrium states at a given
temperature that they satisfy the KMS condition, which is briefly recalled
here. Let $\alpha(t)$, $t \in \RR$, be the
automorphism inducing the time translations on $\fA$ in a chosen 
Lorentz frame. A state $\omega_T$ on $\fA$ satisfies the
KMS condition at temperature $T$ if each correlation function 
\be  \label{e.7.5}
t \mapsto \omega_T(A_1 \alpha(t)(A_2)) \, , \quad A_1, A_2 \in \fA \, ,
\ee
can be extended to the strip 
$S_T \coloneqq \{ z \in \CC : 0 \leq \mbox{Im} z \leq (1/T) \}$, 
is continuous there and analytic in the interior, and has at the upper rim
the boundary value $t \mapsto \omega_T(\alpha(t)(A_2) A_1)$.
Any such state is invariant under the action of the time translations.
The thermodynamic limit states described above comply with this
condition if $\fA$ consists of operators which transform norm-continuously
under the action of the time evolution. It is an open
problem whether this technical assumption can be dropped. 

The discovery of the KMS condition in algebraic
quantum field theory provided a fruitful link with developments
in the theory of operator \mbox{algebras}, \viz modular theory, 
established by Tomita and Takesaki. We briefly summarize here 
some essential points, cf.\ also \cite{Ta,e3}. 
Within the mathematical setting, one deals with
weakly closed operator algebras~$\fR$ which are represented
on some Hilbert space $\cH$ and have a cyclic and separating
vector $\Omega$; that is,  the subspace $\fR \, \Omega$ is dense in
$\cH$ and, apart from~$0$, there is no operator in $\fR$
which annihilates~$\Omega$. Given these ingredients, one
considers the antilinear
operator $S : \fR \, \Omega \rightarrow \cH$ given by
\be  \label{e.7.6}
S \, R \Omega \coloneqq R^* \Omega \, , \quad  R \in \fR \, .
\ee
It is a densely defined, closable operator whose closure has a 
polar decomposition, denoted by $S = J \Delta^{1/2}$.
Here the operator $J$, called modular conjugation,
is an antiunitary involution, $J^2 = 1$; the
operator $\Delta$, called modular operator, is positive and
satisfies $\Delta \Omega = \Omega$.
The central result obtained in this 
setting are the equalities of sets, involving the
adjoint actions of these operators on the algebra $\fR$,  
\be  \label{e.7.7}
J \, \fR \, J = \fR' \, , \qquad \Delta^{it} \, \fR \, \Delta^{-it} = \fR \, ,
\ t \in \RR \, .
\ee
Denoting by $\delta(s)$ the adjoint action of
the unitaries $\Delta^{is}$ on $\fR$, $s \in \RR$, called modular
automorphism group, one considers the correlation functions
\be   \label{e.7.8}
s \mapsto \langle \Omega, R_1 \delta(s)(R_2) \Omega \rangle
\, , \quad R_1, R_2 \in \fR \, .
\ee
Remarkably, they satisfy the KMS condition for temperature $1$; it 
can be replaced by any other value by an adjustment of the exponent  
of the modular operator. 
There is an important converse of this result with
regard to its applications in physics: in the GNS-representation
induced by a KMS state on $\fA$, the dynamics coincides (up to
rescalings) with the corresponding modular automorphism group,
acting on the weak closure of the represented algebra of
observables. 

These observations have found numerous applications in AQFT,
where the occurrence of cyclic and separating vectors for
(sub)algebras of the observables $\fA$ is quite
common (Reeh-Schlieder property \cite{ReSch}). With regard to the physical
interpretation of the theory, a prominent result is
the Bisognano-Wichmann theorem which deals with the 
vacuum representation of the subalgebras of observables localized
in regions bounded by two lightlike planes (wedges).
It says that the modular group for any such wedge algebra and
the vacuum state coincides with the automorphic action of
the one-parameter group of boosts leaving the wedge
invariant. Since the boosts can be interpreted as dynamics
of a uniformly accelerated observer, this result establishes 
the general nature of the Unruh effect in~AQFT \cite{Se}.
Moreover, the respective modular conjugations agree with
the PCT-operator, multiplied with a Poincar\'e transformation
which transforms the parity P into
a reflection about the spatial plane tangent to the wedge. 

The discovery that in this particular case
the modular structure is related to spacetime
symmetries has triggered interest in the 
modular operators of other regions, such as double cones or lightcones.
Indeed, if the underlying theory has a sufficiently big symmetry group,
such as in conformal field theory, the modular groups
of double cone or lightcone algebras
and the vacuum state act like specific subgroups of the conformal
group on these algebras,
cf.\ for example \cite{HiLo}. In general, however, such a
geometric behaviour may not be expected. Even in case of a
non-interacting  massive scalar field the specific properties of the
modular groups for double cones and the vacuum state are not yet known. 

Modular theory has also contributed to progress in the  
structural analysis of AQFT. It is an important consequence
of the Bisognano-Wichmann theorem \cite{BiWi} that a version of
the condition of Haag duality (essential duality) is
satisfied in the vacuum sector. This feature is a  vital ingredient
in sector analysis. Moreover, with the help of modular theory, 
the Murray-von Neumann type was determined of the weak
closures of algebras of observables in various
regions. These algebras were shown to be of type
$\mbox{III}_1$ irrespective of the underlying theory \cite{Fr2}.
This feature is a consequence of the sharp boundaries
of the regions so that observations in the interior and
exterior are strongly entangled.
It results in algebraic properties which 
are quite different from those known of the
algebra of all bounded operators on a Hilbert space.
In particular, there exists no meaningful trace on the
local algebras. This complicates the implementation 
of basic physical concepts, such as entropy. 
One has either to rely on the existence of algebras,
admitting a trace, that contain a given local algebra and which 
are themselves contained in a slightly larger local algebra.  
This corresponds to sealing off a laboratory 
from the outside with walls (split property \cite{DoLo}). 
Or one may enlarge the local algebras by proceeding to the
crossed product with their modular groups. The resulting
von Neumann algebras are known to admit a (not necessarily
unique) trace, they are of type II. It was recently
recognized that these enlargements can be interpreted as
couplings of the local algebras through the modular
groups with the reference systems of observers~\cite{ChLoPeWi,FeJaLoRe}.  

Another area of applications of modular theory is the
analysis of the degrees of freedom of field theoretic
models. The maps introduced in relation \eqref{e.7.1}, based on  
Hamiltonians, can be replaced by similar maps involving
the modular operators. It led to the formulation of
modular nuclearity conditions which are less
restrictive~\cite{BuDaLo}. But they still imply the existence of
product states and projections in the vacuum
sector, cf.\  equation \eqref{e.7.4}. These observations
allowed the construction of a large family of integrable
models within the algebraic framework, which were not
accessible by other constructive means \cite{e4}. Thus the
results of an extensive structural analyses in the
general framework of AQFT provided the ground for these novel
constructive methods. 


\vspace*{-5mm}
\section{Further topics}
\label{sec8}
\setcounter{equation}{0} 
There exist key results which, originally, were
derived in the Wightman framework of quantum field theory, such as   
the spin-statistics theorem and the PCT theorem \cite{StWi}. 
Using the Doplicher-Haag-Roberts theory of
superselection sectors and its later developments, 
these results were established in AQFT in much greater
generality \cite{DoHaRo1,DoHaRo2,BuFr1}. 
For example, the Bose-Fermi alternative of statistics was
derived and not simply assumed, such as in the Wightman framework.  
Moreover, the spin-statistics theorem was established
in massive theories for particles carrying a non-localizable
charge, which do not fit into the Wightman framework.
In all of these theories, the existence of charge conjugate sectors
was established, along with their finite statistics.
In contrast, these features had to
be postulated in the Wightman framework. Finally, 
in low dimensional spacetimes the appearance of braid group
statistics leads to modifications of these results which 
were established in AQFT as well \cite{e4,e9}. 

Point-localized quantum fields do not occur explicitly in the
framework of AQFT. But their existence was established 
under appropriate conditions~\cite{FrHe}, restricting,
for example, the size of the phase space of the theory~\cite{Bos1}.
These fields can be recovered from the local 
algebras as operator valued distributions which transform
covariantly under Poincar\'e transformations. Moreover,
there exist algebraic relations between the pointlike fields 
(operator product expansions) that encode characteristic features 
of a theory~\cite{Bos2}.

Such pointlike structures play also a crucial
role in AQFT on globally hyperbolic spacetimes, where spacetime
symmetries are generically absent. So they cannot be used for
the identification of observables in different regions. Instead, 
one relies on a more fundamental principle of general
covariance \cite{BrFrVe}.
In case of symmetric spacetimes, one recovers from this
principle the automorphic action of the symmetries on
observables. For general spacetimes it implies that point
fields, such as the energy momentum tensor, can be used
in order to identify the observables in different regions.
This fact is, for example, an important ingredient in the
perturbative renormalization of quantum field theories
on curved spacetimes \cite{HoWa1,HoWa2}.


\end{document}